\documentclass[10pt]{iopart}

\pdfoutput=1
\usepackage[square,numbers,sort&compress]{natbib}
\usepackage[breaklinks=true,colorlinks=true,linkcolor=blue,urlcolor=blue,citecolor=blue]{hyperref}
\usepackage{latexsym}
\usepackage{graphics}
\usepackage{graphicx}
\usepackage{epsfig}
\usepackage{mathcomp}
\usepackage{amsopn}
\usepackage{amsfonts}
\usepackage{amstext}
\usepackage{amssymb}
\usepackage{amstext}
\usepackage{subfigure}
\usepackage{braket}
\usepackage{iopams}  
\usepackage{wasysym}

\usepackage{epstopdf}
\usepackage{braket}
\begin{document}

\title{Soliton Dimer-soliton scattering in coupled Quasi-one-dimensional Dipolar Bose-Einstein Condensates}
\author{Gautam Hegde, Pranay Nayak, Ratheejit Ghosh and Rejish Nath}
\address{Department of Physics, Indian Institute of Science Education and Research, Dr. Homi Bhabha Road, Pune 411008, Maharashtra, India.}
\vspace{10pt}

\begin{abstract}
We discuss scattering between a bright soliton and a soliton dimer in coupled quasi-one-dimensional dipolar Bose-Einstein condensates. The dimer is formed by each soliton from both tubes due to the attractive inter-layer dipole-dipole interaction. The dipoles within each tube repel each other, and a stable, bright soliton is stabilized via attractive contact interactions. In general, the scattering is inelastic, transferring the kinetic energy into internal modes of both soliton dimer and single soliton. Our studies reveal rich scattering scenarios, including dimer-soliton repulsion at small initial velocities, exchange of atoms between dimer and single soliton and soliton fusion at intermediate velocities. Interestingly, for some particular initial velocities, the dimer-soliton scattering results in a state of two dimers. At large initial velocities, the scattering is elastic as expected.
\end{abstract}

%
%
\submitto{\JPB}
%
\maketitle
%
\ioptwocol
\tableofcontents
\section{Introduction}
Collisional or scattering dynamics of solitons is of interest in both fundamental science and applications \cite{has95, mol06, kev08}. True solitons in one dimension, owing to the integrability of the corresponding non-linear equation, should pass through each other without suffering any changes in shape and velocity, i.e., undergoes an elastic scattering. The bright solitons typically interact upon contact, with an effective interaction generated by the interference of the two wave packets, with its range decays exponentially \cite{gor83, mit87}. The nature of soliton interaction, whether being attractive or repulsive, depends on their relative phase as well. 

Bose-Einstein condensates (BECs) opened up a new avenue to explore the soliton physics in a more controlled manner in particular for bright solitons \cite{kev08, ngu14, kha02a, str02, mar13, med14, kha02, cor06, ngu17, eve17}. In condensates, strictly speaking, we do not deal with solitons but solitary waves due to both the external confinement and their quasi-one-dimensional (Q1D) nature and thus can experience in-elastic collisions \cite{mar08}. The collisional dynamics between bright solitons is studied in condensates with both contact interactions \cite{mar08,ngu14, ngu17,mar07, ger13, bil13, mcd14, lep16, mez19, wal20,par08, xut18, bil11, cho20, kha11, mar12} and dipole-dipole interactions (DDIs) \cite{nat07, abd12,bai15, chi14, cue09, eic12, gao21, abd14, you11,edm17, abd13}, which have been exploited to generate entangled soliton pairs \cite{ger13, bil13} and design interferometers \cite{ mcd14, wal20,mar12, pol13, hel15}. Contrary to the short-range condensates, in dipolar BECs, the long range and anisotropic nature of DDI lead to novel scenarios such as the multi-dimensional solitons \cite{rag15, nat08, mis16, ped05, tik08, che17} and interlayer effects \cite{hua10,kla09, kob09, lak12c, lak12b,mul11}. The latter include the stabilization of soliton complexes such as molecules or dimers, crystals and filaments \cite{bai15,lak12c,lak12a,elh19}.

This paper analyses the scattering dynamics of three bright solitons realized in two coupled Q1D traps. Two out of three solitons from each tube form a dimer and collide with a single soliton. To appreciate the new dynamics emerging in the dimer-soliton setup, we first briefly discuss scattering two Q1D solitons in a single tube. In the latter case, we observe the repulsion between solitons for small initial velocities, both inelastic destruction and fusion at intermediate velocities and elastic scattering at large initial velocities.  
The inelastic scattering is due to the transfer of the kinetic energy into the internal modes of the solitons. The fusion and destruction of solitons indicate resonant scattering where the kinetic energy is completely transferred to the internal degrees of freedom. The dimer-soliton collision reveals more inelastic resonances as a function of the initial velocity, as they possess more internal modes. A complete characterization of the role of internal modes is not possible due to their dynamical nature. Here, we explore the general scattering scenarios in the dimer-soliton setup for fixed interaction strengths. A different set of interaction parameters would still provide qualitatively similar dynamics but shows a different dependence on the initial velocity. Our studies reveal that for small initial velocities, the dimer and soliton repel each other, whereas, for intermediate velocities, an exchange of atoms between dimer and soliton occurs, which lead to non-trivial inelastic scattering dynamics. One such scenario is the fusion of solitons in the upper layer, transferring the kinetic energy entirely to the internal modes, and the impact can also destroy the soliton in the bottom layer. Among the different scattering scenarios we discuss, the most intriguing is the emergence of two soliton dimers after the collision. As expected, there is an elastic scattering between the dimer and the soliton at large initial velocities.

The paper is structured as follows. In section~\ref{sup}, we introduce the physical setup and the governing equations. In section~\ref{2ss}, we briefly discuss the scattering dynamics of two Q1D solitons in a single tube. In section~\ref{dss}, we discuss the scattering of a soliton-dimer and a single soliton realized in two tubes. Finally, we conclude with an outlook in section~\ref{sum}.

\section{Setup and Model}
\label{sup}
\begin{figure}
\centering
\includegraphics[width=1. \columnwidth]{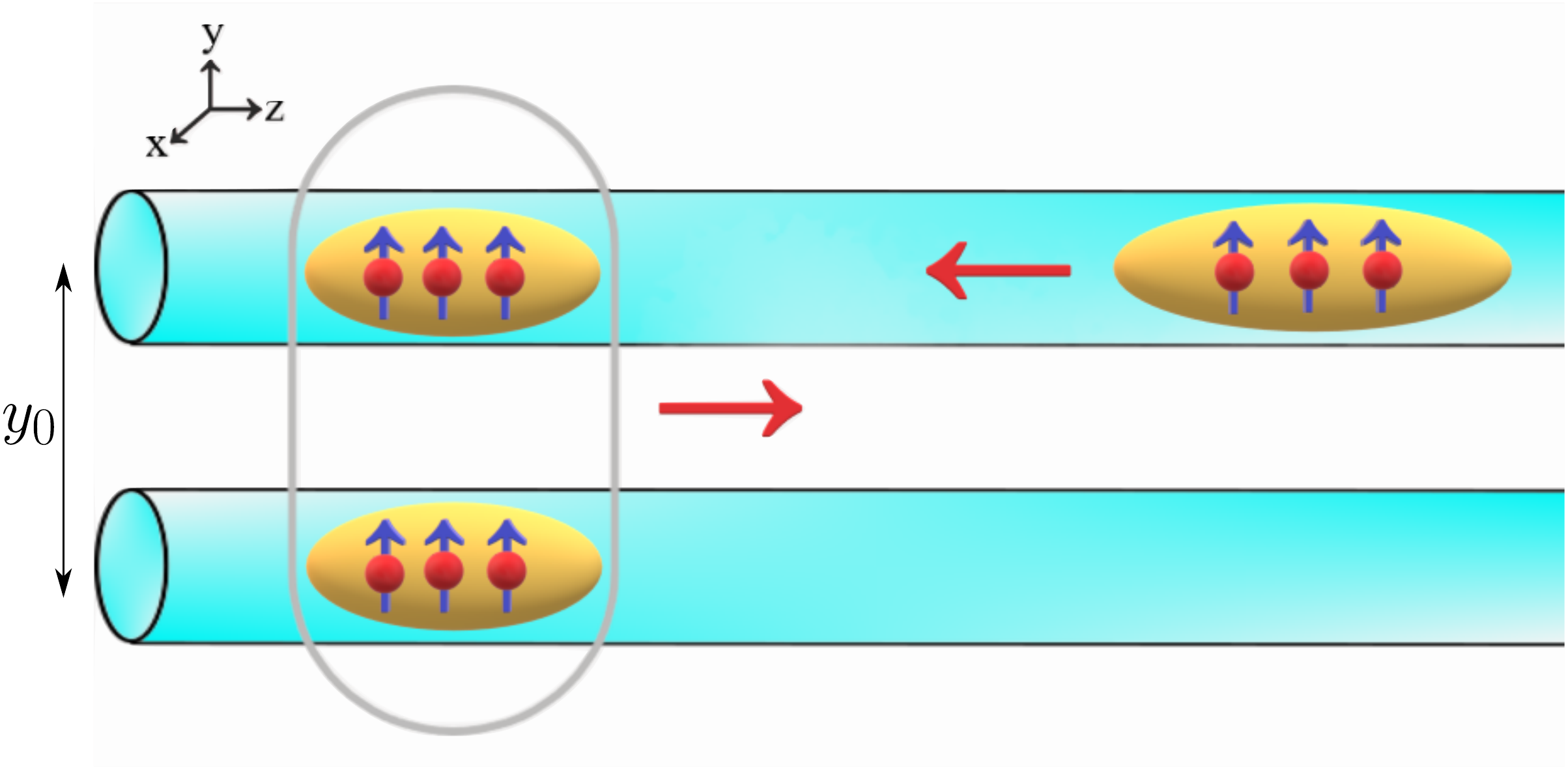}
\caption{Schematic diagram of the scattering setup of couples Q1D dipolar BECs. The tubes are separated along the $y$-axis by a dstance of $y_0$. The two solitons on the left side form a dimer due to the attractive DDI. Dipoles are oriented along the $y$-axis and therefore the solitons in the top layer repel each other.}
\label{fig:1}
\end{figure}
We consider a setup of coupled Q1D dipolar BECs in which the radial direction is harmonically confined with a frequency $\omega_\perp$ and no confinement along the $z$ direction as shown in figure~\ref{fig:1}. The two tubes are separated along the $y$-axis by a separation of $y_0$. We assume the dipoles are aligned along the $y$-axis, and both $\omega_\perp$ and $y_0$ are sufficiently large to ensure no hopping of particles between the tubes. The dipoles attract along the $y$-axis while repelling along the longitudinal $z$-axis. Due to the attractive inter-layer DDIs (along $y$-axis), a soliton dimer can be formed using one soliton from each layer \cite{lak12c}. Effectively, we have a soliton dimer shared between the two tubes and a single soliton, which is well separated from the dimer along the z-axis. The condensates in each tube are described by coupled non-linear Gross-Pitaevskii equations (NLGPE)
\begin{eqnarray}
\label{GPE}
i \hbar \partial_{t} \Psi_{j}(\mathbf{r}) &= \left[-\frac{\hbar^{2}}{2 m} \nabla^{2}+U_{j}(\mathbf{r})+g \left|\Psi_{j}(\mathbf{r})\right|^{2}\right.
\\
& \left.+\sum_{m=1}^{2} \int d \mathbf{r}^{\prime} V_{d}\left(\mathbf{r}-\mathbf{r}^{\prime}\right)\left|\Psi_{m}\left(\mathbf{r}^{\prime}\right)\right|^{2}\right] \Psi_{j}(\mathbf{r}, t), \nonumber
\end{eqnarray}
where $\Psi_{j}$ is the wavefunction of the condensate in the $j$th layer, $U_{1}(\mathbf{r})=m \omega_{\perp}^{2}[x^{2}+\left(y-y_0\right)^2]/2$ and $U_{2}(\mathbf{r})=m \omega_{\perp}^{2}(x^{2}+y^{2})/2$ are the external trapping potentials of first and second tubes, respectively. $V_d(\textbf{r})=g_d(1-3\cos^2\theta)/|\textbf{r}|^3$ is the dipole-dipole potential with the interaction strength $g_d=N\mu_{0} \mu^{2} / 4 \pi$, with $\mu$ being the dipole moment and $\theta$ being the angle between dipole axis $y$ and radial vector $\textbf{r}$ between two dipoles. $N$ is the number of atoms per soliton. Hence the upper tube has $2N$ atoms, and the lower one has $N$ atoms. The parameter $g=4 \pi a\hbar^{2}N/m$ characterizes the strength of the contact interaction with $a$ being the $s$-wave scattering length. The wavefunctions satisfy the normalization conditions $\int d^3r|\Psi_1|^2=2$ and $\int d^3r|\Psi_2|^2=1$, leaving all three solitons with same number of atoms at $t=0$.  The condensates are strongly confined in the $xy$-plane, and the chemical potential of each condensate is much smaller than $\hbar\omega_\perp$ to ensure the Q1D nature of the condensates. In that  case, we can factorize the BEC wave function at each tube as $\Psi_j(x,y,z)=\psi_j(z)\phi_j(x,y)$, where $\phi_j(x,y)$ is the ground-state wave function of the $xy$ harmonic oscillator, which is a Gaussian function. Using the Fourier transform of the dipolar potential, convolution theorem and integrating over the radial directions, we arrive at coupled 1D NLGPEs \cite{lak12a,lak12b,lak12c,cai10},
\begin{eqnarray}
\label{2}
i \hbar \partial_{t} \psi_{j}(z)=\left[-\frac{\hbar^{2}}{2 m} \partial_{z}^{2}+\frac{g}{2 \pi l_{\perp}^{2}} |\psi_{j}(z)|^2\right.\nonumber  \\ \left.
+\frac{2g_{d}}{3} \sum_{m=1}^{M} \frac{1}{2 \pi}\int d k_{z} e^{i k_{z} z} n_{m}\left(k_{z}\right) F_{mj}\left(k_{z}\right)\right] \psi_{j}(z),
\end{eqnarray}
where $\hat{n}_m(k_z)$ is the Fourier transform of density $|\psi_m(z)|^2$, and 
\begin{eqnarray}
    F_{ij}\left(k_{z}\right)&=&\int \int d k_{x} d k_{y}\left(\frac{3 k_{y}^{2}}{k_{x}^{2}+k_{y}^{2}+k_{z}^{2}}-1\right)  \nonumber \\
    &&\times  e^{-\frac{1}{2}\left(k_{x}^{2}+k_{y}^{2}\right) l_{\perp}^{2}-\imath k_{y} (y_i - y_j)}. 
\end{eqnarray}
We use the dimensionless quantities $\tilde{g}= g/2 \pi \hbar \omega _\perp l_\perp^3$ and $\tilde g_d= g_d/2 \pi \hbar \omega _\perp l_\perp^3$ with $l_\perp = \sqrt{\hbar/m\omega_\perp}$. Also, we take $y_0=7.5l_\perp$ throughout the paper. The stable/unstable regions of both soliton dimer and a single Q1D bright soliton as a function of $\tilde{g}$ and $\tilde{g}_d$ for the configuration we consider are discussed in \cite{lak12c}. Based on that, we obtain the ground states of a soliton dimer and a single soliton via imaginary time evolution of the Q1D NLGPEs for $\tilde{g}=-0.9,\tilde{g}_d=0.37$, which we kept through out the paper. We want to stress that, a different set of values for $\tilde{g}$ and $\tilde{g}_d$, in the soliton/dimer regime, as we verified, would give us qualitatively the same results that we show below. Additionally, for the parameters in the soliton regime, in dipolar BECs, there will be an energy correction or the Lee-Huang-Yang (LHY) correction arising from the quantum fluctuations. The LHY correction stabilizes a collapsing three-dimensional dipolar BEC into quantum droplets, especially at higher densities \cite{wac16}. In the Q1D regime, there is no collapse, and the LHY correction only modifies the soliton density slightly \cite{edl17, pri21} and can be safely ignored in our studies. And further, the LHY correction is usually less significant for atoms with smaller dipole moments (e.g., chromium atoms) and low densities. For a chromium condensate, $\tilde{g}=-0.9,\tilde{g}_d=0.37$ corresponds to a BEC of $N=1000$, $\omega_\perp=2\pi\times 182$ Hz and $a=-9a_0$, where $a_0$ is the Bohr radius.

Before indulging in the dynamics of soliton dimer-soliton scattering, we briefly outline the main scenarios in the scattering of two Q1D solitons in a single tube. The scattering of two Q1D solitons in disconnected parallel tubes are discussed in \cite{nat07} and is characterized by inelastic scattering resonances in which the kinetic energy is resonantly transferred into the internal modes of the soliton for an intermediate range of initial velocities. Such resonant inelastic scattering may destroy the solitons. 

\section{Soliton-soliton scattering in a single tube}
\label{2ss}
\begin{figure}
\centering
\includegraphics[width=1. \columnwidth]{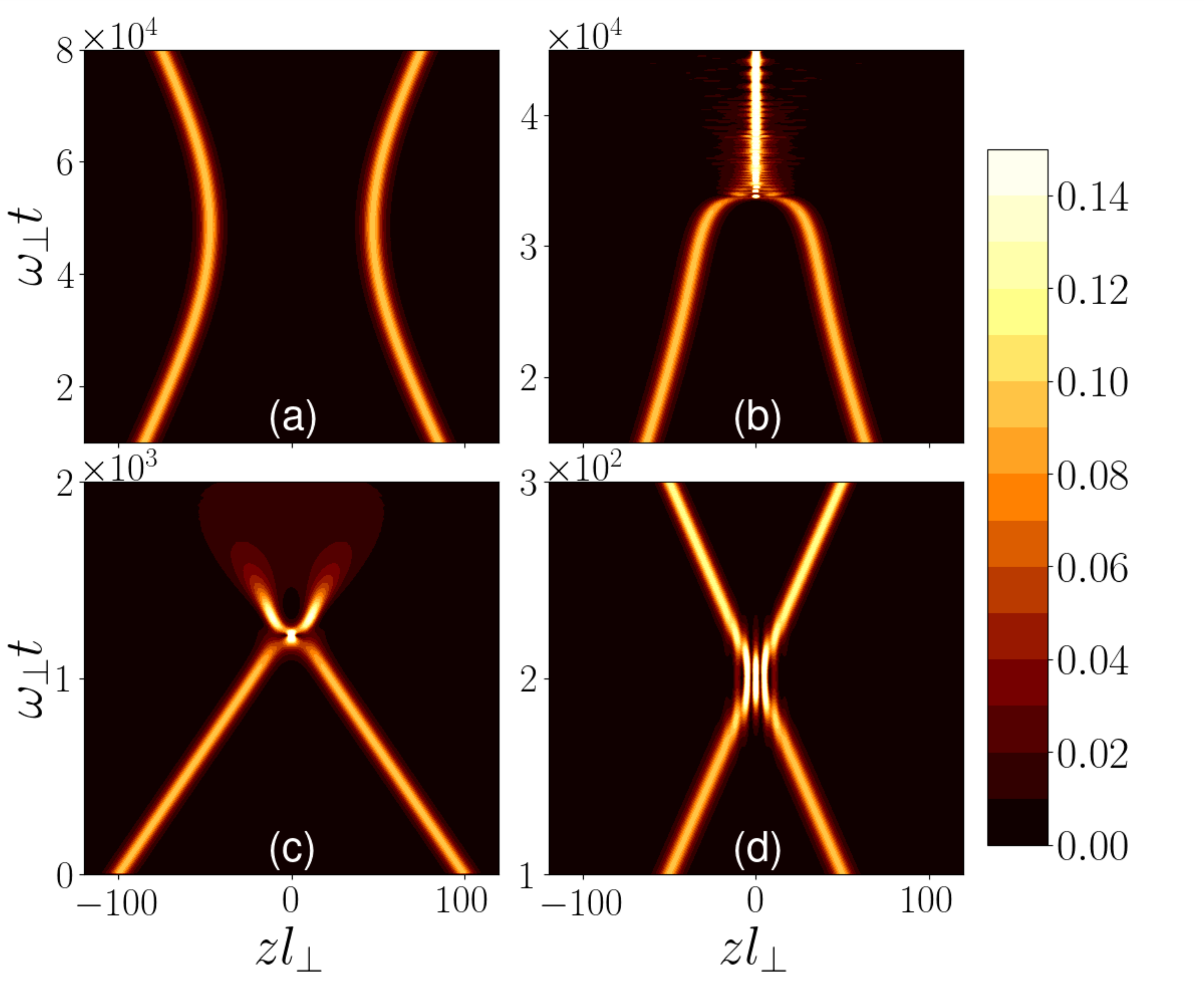}
\caption{Different scenarios in two soliton scattering dynamics for $\tilde{g}=-0.9,\tilde{g}_d=0.37$. We are showing the soliton trajectories along the $z$-axis as a function of time. (a) depicts the repulsion between the solitons for a small initial momentum of $k_i l_\perp = 0.0015$. (b) shows the solitons fusion at $k_i l_\perp = 0.0025$ whereas (c) shows the soliton destruction at $k_i l_\perp = 0.078$. (d) shows the elastic scattering at a large initial momentum of $k_i l_\perp = 0.5$.}
\label{fig:2}
\end{figure}
Here, we briefly discuss the scattering of two Q1D solitons in a single tube along the $z$-axis. As mentioned before, we take $\tilde g=-0.9$ (attractive short-range interaction) and $\tilde g_d=0.37$. Since the dipoles are oriented along the $y$-axis, the solitons repel each other along the $z$-axis. Note that the contact interactions are kept attractive to stabilize the initial solitons. We take the initial positions of the solitons are at $z(t=0)=\pm 100 l_\perp$, which is sufficiently far enough to neglect the DDI between them. Then, each of the solitons is provided with equal initial velocities but in opposite directions. We observe four scattering scenarios as a function of the initial velocity or equivalently initial momentum $k_i$. For sufficiently small initial velocities, the kinetic energy of the solitons is insufficient to overcome the dipolar repulsion, and the solitons repel away from each other, as shown in figure~\ref{fig:2}(a). As the initial velocity increases, the solitons come closer and closer, and above a critical initial speed, they come in contact. For the intermediate velocities, we observe both soliton fusion and soliton destruction. In the case of soliton fusion [see figure~\ref{fig:2}(b)], the resultant final soliton remains at rest at the point of contact, with a partial excitation of its breathing mode. In the destructive case [see figure~\ref{fig:2}(c)], the kinetic energy is mostly transferred to the internal modes, and the solitons disperse (undergoes expansion) after emerging out of the collision. At sufficiently large velocities, they undergo elastic scattering similar to that of an ideal soliton in which the solitons pass through each other. The latter scenario is also characterized by an interference pattern in the contact region, as seen in figure.~\ref{fig:2}(d). The initial velocities at which the four scattering scenarios occur depends critically on the interaction strengths $\tilde g$ and $\tilde g_d$.

\section{Dimer-soliton scattering}
\label{dss}

\begin{figure}
\centering
\includegraphics[width=1.\columnwidth]{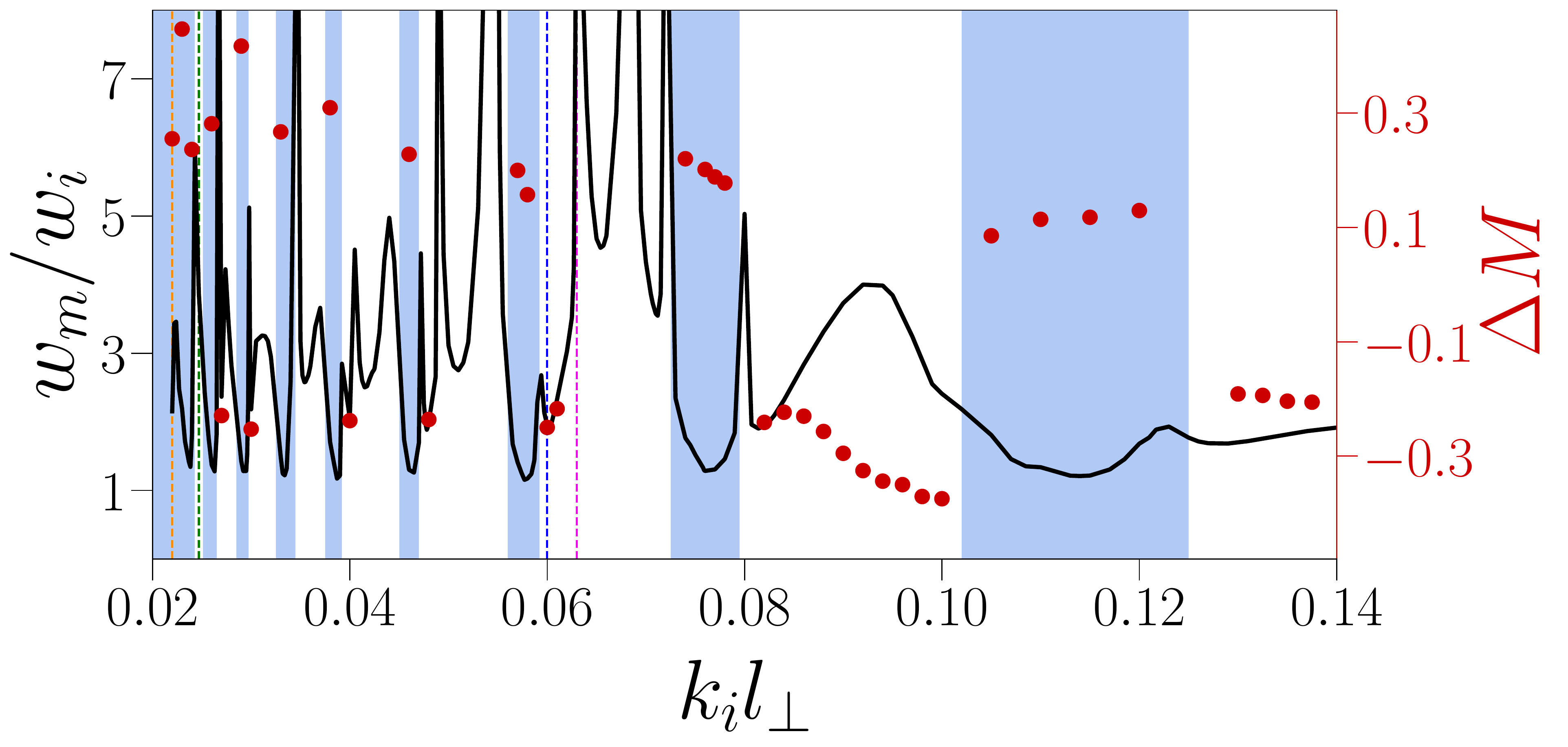}
\caption{The ratio of maximum width $w_m$ and the initial width $w_i$ of the soliton in the second layer as a function of the initial momentum $k_i$. Here, the value of $k_i$ is restricted to the intermediate values for which inelastic scattering occurs. We also show the mass imbalance $\Delta M$ of the first tube calculated at a sufficiently long time after the scattering.}
\label{fig:3}
\end{figure}

In this section, we study the scattering between a soliton dimer and a single soliton for $\tilde{g}=-0.9,\tilde{g}_d=0.37$ in a coupled Q1d dipolar BECs. At $t=0$, the ground state solutions of a soliton dimer and a single soliton are placed at $z=\pm 100l_\perp$. Then they are subjected to collisions providing identical initial velocities to both the dimer and the single soliton. The setup is such that we have two solitons in the first tube and a single soliton in the second tube. Before the collision, the lone soliton in the second tube forms a dimer with the soliton in the first tube. In general, the soliton collision is sensitive to the phase difference between them. Thus, the initial separation can play an essential role due to the long-range nature of DDI. By the time the solitons collide physically, the inter DDI induces different phases in three solitons.

\begin{figure*}
\centering
\includegraphics[width=1.6\columnwidth]{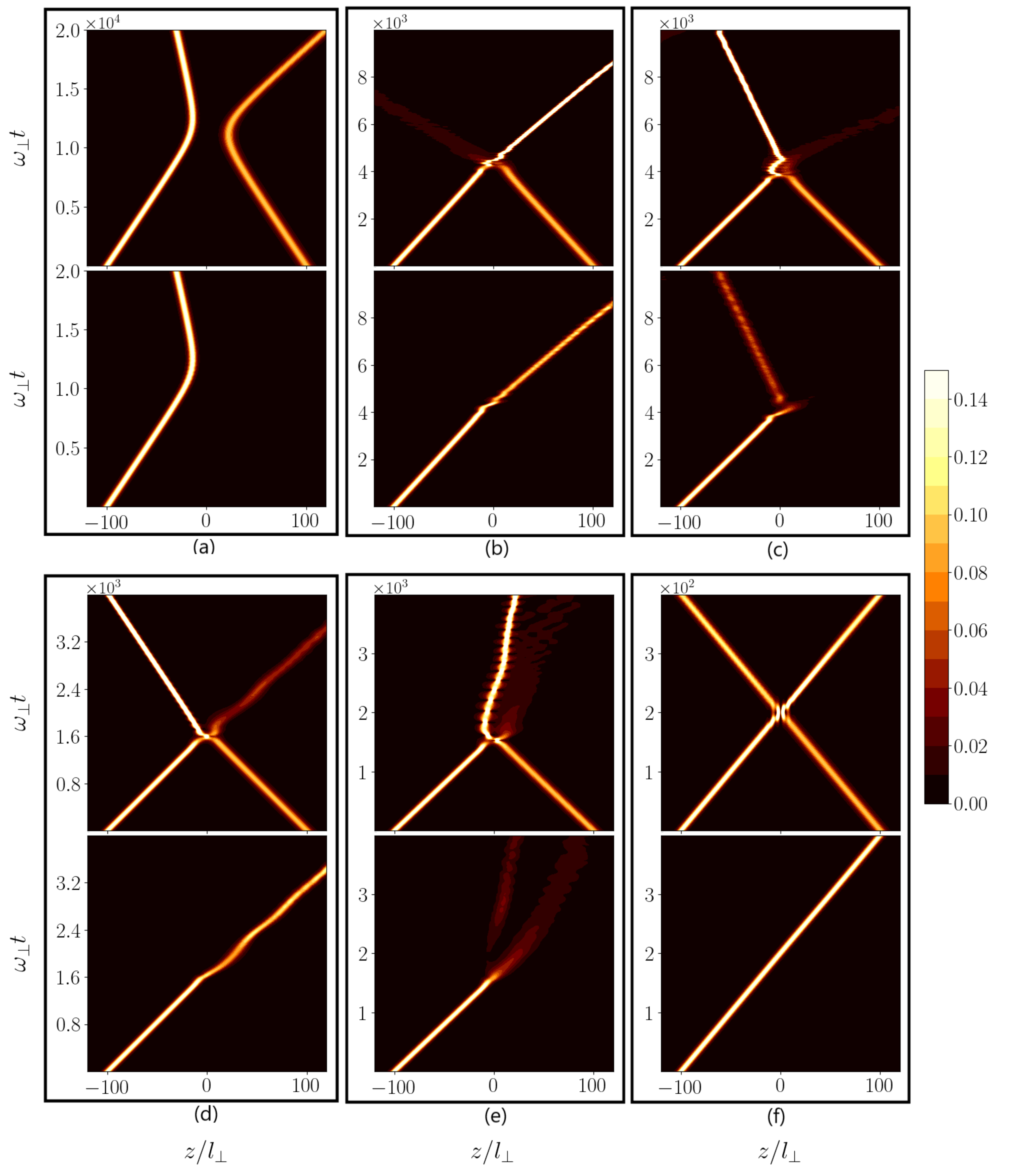}
\caption{Different scenarios in the scattering between soliton dimer and a single soliton for $\tilde{g}=-0.9,\tilde{g}_d=0.37$. We are showing the soliton trajectories in both tubes along the $z$-axis as a function of time. In each box, the upper figure depicts the two solitons in the first tube and the bottom figure for the lone soliton in the second tube. (a) depicts the repulsion between the solitons for a small initial momentum of $k_i l_\perp = 0.008$. (b)-(e) show the inelastic scattering at the intermediate velocities which includes the soliton fusion in the first tube and soliton destruction in the second tube. The initial velocities are (b) $k_i l_\perp = 0.022$, (c) $k_i l_\perp = 0.0247$, (d) $k_i l_\perp = 0.06$ and (e) $k_i l_\perp = 0.063$. (e) depicts the soliton fusion in the upper layer and (f) is for large initial velocity of $k_i l_\perp = 0.5$ for which the elastic scattering occurs.}
\label{fig:4}
\end{figure*}

As we show below, the ratio $w_m/w_i$ together with the mass imbalance in the first tube can provide us with a hint at the nature of the scattering, where $w_m$ is the maximum width attained by the lone soliton in the second tube after the collision and $w_i$ is its initial width. A large $w_m$ indicates a significant transfer of kinetic energy into the internal modes and if that happens resonantly, it can even blow up the soliton ($w_m\to\infty$) or destruct abruptly. The mass imbalance is defined as the normalized difference in probabilities or masses in the left and right halves of the upper tube: $M_l=\int_0^{L/2}|\psi_1|^2dz$ and $M_r=\int_{L/2}^L|\psi_1|^2dz$ at a sufficiently long time after the collision, where $L$ is the total box size, i.e., 
\begin{equation}
\Delta M=\frac{M_r-M_l}{M_r+M_l},
\end{equation}
where $M_l+M_r=2$ in our case. $\Delta M<0$ ($\Delta M>0$) implies there is more fraction of condensates in the left (right) half of the cylindrical tube after the scattering. The behaviour of $w_m$ as a function of $k_i$ for $k_il_\perp\geq 0.02$ is shown in  figure~\ref{fig:3}. The different scattering scenarios are shown in figure~\ref{fig:4} and the corresponding $k_i$ are marked by dashed vertical lines in figure~\ref{fig:3}. Keep in mind that the $k_i$'s at which a particular scattering scenario occurs depends critically on the values of $\tilde g$ and $\tilde g_d$.

For small initial velocities (corresponds to $k_il_\perp<0.02$), both dimer and single soliton reflect each other as shown in figure~\ref{fig:4}(a) for $k_il_\perp=0.008$, which is also identical to that of two-soliton scattering discussed in section~\ref{2ss}. The latter occurs due to the repulsive interaction between the solitons in the upper tube. In figure~\ref{fig:4}(a), the upper plot shows the trajectories of two solitons in the first tube, and the lower plot shows the trajectory of the soliton in the second tube. For the scattering scenario in figure~\ref{fig:4}(a), we have $\Delta M=0$ (not shown in figure~\ref{fig:3}) since there is no transfer of atoms between the solitons in the upper layer.

Upon increasing the initial speeds further, enough to overcome the long-range repulsion, we observe non-trivial collisional dynamics as a function of the initial velocity that is entirely different from simple two soliton scattering. Due to more degrees of freedom, the dimer-soliton setup possesses more low-lying modes than the two-soliton system, which intuitively explains why one would expect more complicated scattering dynamics in the former. That is also supported by the series of inelastic resonance peaks in the $w_m/w_i$ vs $k_i$ plot shown in figure~\ref{fig:3}. Sharp and varying widths of resonance peaks indicate the non-trivial dependence on the initial velocities. In both two soliton and three soliton cases, the modes dynamically change as the solitons approach each other. The latter prevents a comprehensive analysis of the role of these elementary modes on the soliton scattering. 

In figure~\ref{fig:4}(b), for $k_il_\perp=0.022$, after the scattering, the dimer passes through the single soliton, but destructing the single soliton which it collided with [see the upper panel in figure~\ref{fig:4}(b)]. The lower panel in figure~\ref{fig:4}(b) indicates that the breathing mode of the lone soliton in the second tube is excited by the collision event. It signifies the importance of inter-layer effects in disconnected dipolar condensates. In addition, for this case, we see that $\Delta M>0$ (figure~\ref{fig:3}) implying that there are more atoms on the right half of the first tube after the collision. That means there was a transfer of atoms among the solitons in the first layer, leaving a bigger soliton in the dimer, making it asymmetric after the collision. All the blue-shaded regions in figure~\ref{fig:3} correspond to the same dynamics as that for $k_il_\perp=0.022$, indicating the non-monotonous dependence of scattering dynamics on the initial velocity. Also, the density plot in figure~\ref{fig:4}(b) reveals that the velocity of the dimer increased slightly after the destructive collision with the single soliton. 

Due to the sharp resonant peaks at the intermediate initial velocities, a slight change in $k_i$ can lead to drastically different dynamics as shown in figure~\ref{fig:4}(c) for which $k_il_\perp=0.0247$. After the scattering, a dimer does survive but moves in the opposite direction. The impact of the scattering destroys the remaining soliton in the first tube. As one can see, the scattering significantly affects the soliton in the second tube due to the DDI between the tubes. We also see that figure~\ref{fig:4}(c) is almost a mirror image of what is depicted in figure~\ref{fig:4}(b). Figure~\ref{fig:4}(c) is characterized by $\Delta M<0$ since, in the upper tube, a significant population is moved towards the left. At another value of $k_i$ [see figure~\ref{fig:4}(d)], we see another prominent scenario, where all three solitons survive after the scattering, leaving a new dimer and a single soliton. Two features are visible in this case: (i) in the first tube, the final single soliton carries more atoms (hence higher density) than the initial single soliton and (ii) the vibrational modes of the dimer are also weakly excited after the collision.

Another interesting scenario is the soliton fusion in the first tube, which is shown in figure~\ref{fig:4}(e) for $k_il_\perp=0.063$. After the collision, almost a complete transfer of kinetic energy into the internal modes leaves the two solitons in the first tube fused. It also results in the excitation of the soliton in the lower tube, which can lead to its destruction depending on the initial kinetic energy, as is the case here. It is clear from the top panel of figure~\ref{fig:4}(e) that the final soliton in the first tube exhibits strong breathing mode oscillations, and also a jet of atoms keep emitting from it. The destruction of the soliton in the bottom tube is featured by two structureless clouds of atoms [see the bottom panel in figure~\ref{fig:4}(e)]. At high initial velocities, shown in figure~\ref{fig:4}(f), the dimer and the soliton pass through each other after the collision indicating an elastic scattering.

\begin{figure}
\centering
\includegraphics[width=1.0\columnwidth]{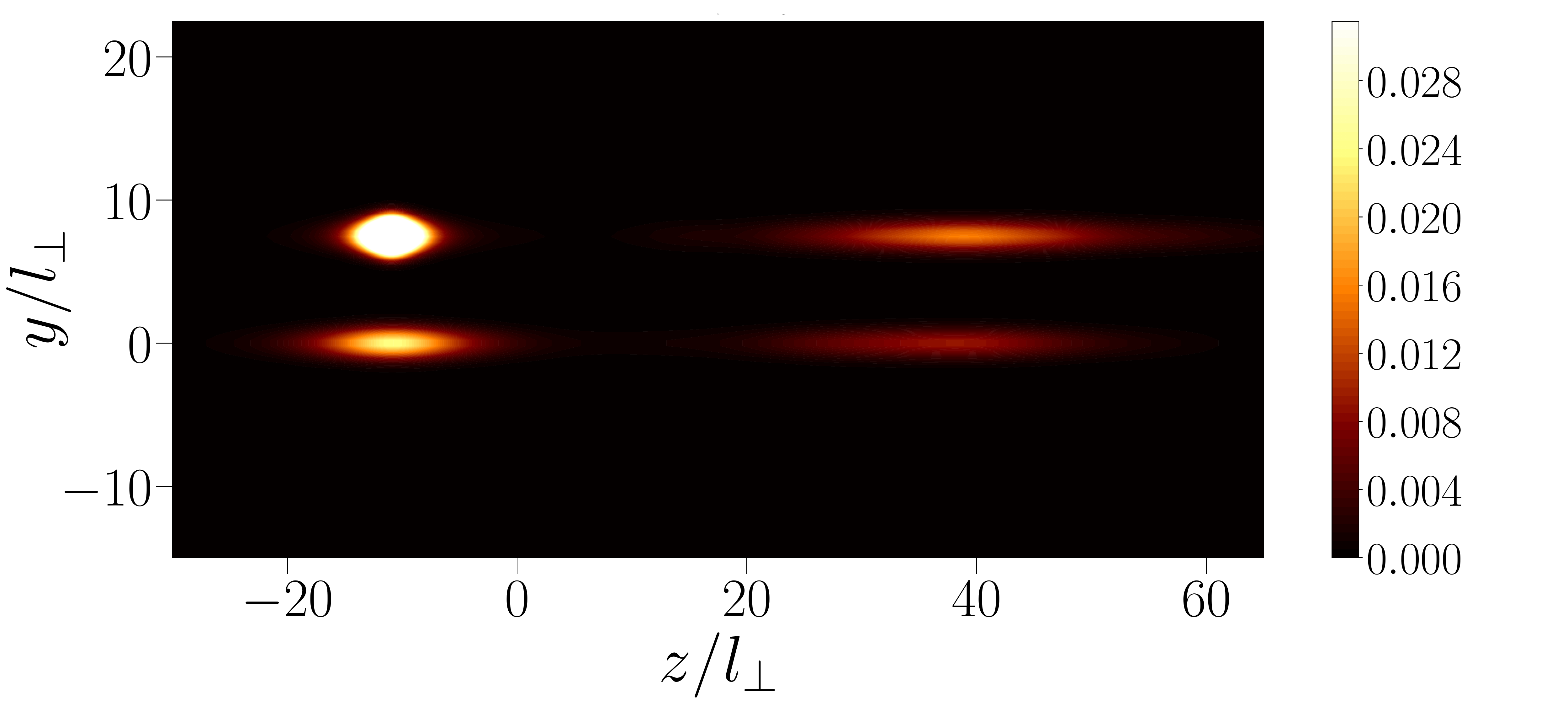}
\caption{The density of the condensates in both layers after the scattering for $k_il_\perp=0.02671$ at $\omega_\perp t=4680$, for which we see the emergence of two dimers. The values of other parameters are the same as that of figure~\ref{fig:4}.}
\label{fig:5}
\end{figure}

The foremost exciting scenario we observed is the emergence of an additional dimer (effectively four solitons) from the dimer-soliton (three solitons) scattering. An example is shown in figure~\ref{fig:5}, in which two soliton dimers are moving away from each other after the collision. It indicates that the soliton in the bottom tube breaks up into two solitons upon collision. Due to the asymmetry of the setup, it is hard to obtain the case in which the soliton in the lower tube breaks up into two identical solitons. We would also like to stress that scattering dynamics in multilayer dipolar condensates generally show high sensitivity to initial separations and velocities apart from their dependence on interaction strengths.
\section{Summary and Outlook}
\label{sum}
In conclusion, we studied the scattering dynamics between a soliton dimer and a single soliton in a setup of coupled Q1D dipolar condensates. The long-range and anisotropic nature leads to interesting inelastic scattering dynamics, and in particular, the effect of interlayer DDI becomes manifest in all scattering scenarios. The intriguing nature of scattering dynamics is attributed to the internal modes of the soliton and dimer. Our studies also open up many possibilities to explore shortly, for instance, the role of the orientation of dipoles in coupled condensates and the potential to engineer soliton interferometers using dipolar solitons. Another exciting aspect is to explore the role of the initial relative phase between the soliton dimer and the single soliton. The latter in the presence of long-range DDI may lead to even more complex scattering dynamics. Indeed, more complex collisional dynamics is expected with more solitons, for example, for the case of four or more dipolar Q1D solitons.

\section{Acknowledgements}
R.N. acknowledges DST-SERB for Swarnajayanti fellowship File No. SB/SJF/2020-21/19. We acknowledge National Supercomputing Mission (NSM) for providing computing resources of 'PARAM Brahma' at IISER Pune, which is implemented by C-DAC and supported by the Ministry of Electronics and Information Technology (MeitY) and Department of Science and Technology (DST), Government of India. G.H. and P.N. acknowledge the funding from DST India through an INSPIRE scholarship.

\providecommand{\newblock}{}


\begin{thebibliography}{10}
\expandafter\ifx\csname url\endcsname\relax
  \def\url#1{{\tt #1}}\fi
\expandafter\ifx\csname urlprefix\endcsname\relax\def\urlprefix{URL }\fi
\providecommand{\eprint}[2][]{\url{#2}}

\bibitem{has95}Hasegawa A and Kodama Y 1995 {\em Solitons in Optical Communications}
Clarendon Press, Oxford, UK
\bibitem{mol06} Mollenauer L F and Gordon J P 2006, {\em Solitons in Optical Fibers:
Fundamentals and Applications} Academic Press, San Diego, CA
\bibitem{kev08} Kevrekidis P G, Frantzeskakis D J, and Carretero-Gonz\'alez R 2007 {\em Emergent Nonlinear Phenomena in Bose-Einstein Condensates: Theory and Experiment} (Berlin: Springer)
\bibitem{gor83}Gordon J P 1983  {\em Opt. Lett.} {\bf 8} 596
\bibitem{mit87}Mitschke F M and Mollenauer L F 1987 {\em Opt. Lett.} {\bf 12} 355
\bibitem{ngu14}Nguyen J H V, Dyke P, Luo D, Malomed B A and Hulet R G 2014 {\em Nat.Phys.} {\bf 10} 918
\bibitem{kha02a}Khaykovich L, Schreck F, Ferrai G, Bourdel T, Cubizolles J, Carr LD, Castin Y and Salomon C 2002 {\em Science} {\bf 296} 1290
\bibitem{str02}Strecker K E, Partridge G B, Truscott A G and Hulet R G 2002 {\em Nature} {\bf 417} 150
\bibitem{mar13}Marchant A L, Billam T P, Wiles T P, Yu M M H, Gardiner S A and Cornish S L 2013 {\em Nat. Commun.} {\bf 4} 1865
\bibitem{med14}Medley P, Minar M A, Cizek N C, Berryrieser D and Kasevich M A 2014 {\em Phys. Rev. Lett.} {\bf 112}
060401
\bibitem{kha02}Khawaja U A, Stoof H T C, Hulet R G, Strecker K E and Partridge G B 2002 {\em Phys. Rev. Lett.} {\bf 89} 200404
\bibitem{cor06}Cornish S L, Thompson S T and Wieman C E 2006 {\em Phys. Rev. Lett.} {\bf 96} 170401
\bibitem{ngu17}Nguyen J H V, LuoD and Hulet R G 2017 {\em Science} {\bf 356} 422
\bibitem{eve17} Everitt P J et al 2017 {\em Phys. Rev. A} {\bf 96} 041601(R)
\bibitem{mar08}Martin A D, Adams C S and Gardiner S A 2008 {\em Phys. Rev. A} {\bf 77} 013620
\bibitem{mar07}Martin A D, Adams C S and Gardiner S A 2007 {\em Phys. Rev. Lett.} {\bf 98} 020402
\bibitem{ger13}Gertjerenken B, Billam T P, Blackley C L, Sueur C R L, Khaykovich L, Cornish S L and Weiss C 2013 {\em Phys. Rev. Lett.} {\bf 111} 100406
\bibitem{bil13}Billam T P, Blackley C L, Gertjerenken B, Cornish S L and Weiss C 2013 {\em J. Phys.: Conf. Ser.} {\bf 497} 012033
\bibitem{mcd14} McDonald G D et al 2014 {\em Phys. Rev. Lett.} {\bf 113} 013002
\bibitem{lep16}Lepoutre S et al 2016 {\em Phys. Rev. A} {\bf 94} 053626
\bibitem{mez19}  Me\ifmmode \check{z}\else \v{z}\fi{}nar\ifmmode \check{s}\else \v{s}\fi{}i\ifmmode \check{c}\else \v{c}\fi{} T et al 2019 {\em Phys. Rev. A} {\bf 99} 033625
\bibitem{wal20}Wales O J et al 2020 {\em Commun. Phys.} {\bf 3} 51
\bibitem{par08}Parker N G, Martin A M, Cornish S L and Adams C S 2008 {\em J. Phys. B At. Mol. Opt. Phys.} {\bf 41} 045303 
\bibitem{xut18} Xu T F and Zhang C 2018 {\em Chaos Solitons Fract.} {\bf 117} 209
\bibitem{bil11}Billam T P, Cornish S L and Gardiner S A 2011 {\em Phys. Rev. A} {\bf 83} 041602 
\bibitem{cho20}Choudhury S, Sreedharan A, Mukherjee R, Streltsov A and W\"uster S 2020 {\em Phys. Rev. A} {\bf 101}, 043604
\bibitem{kha11} Khawaja U Al and  Stoof H T C 2011 {\em New J. Phys.} {\bf 13} 085003
\bibitem{mar12} Martin A D and Ruostekoski J 2012 {\em New J. Phys.} {\bf 14} 043040 
\bibitem{nat07} Nath R, Pedri P and Santos L 2007 {\em Phys. Rev. A} {\bf 76} 013606
\bibitem{nat09} Nath R, Pedri P and Santos L 2009 {\em Phys. Rev. Lett.} {\bf 102} 050401
\bibitem{abd12}Abdullaev F K and Brazhnyi V A 2012 {\em J. Phys. B: At. Mol. Opt. Phys.} {\bf 45}, 085301
\bibitem{bai15}Baizakov B B, Al-Marzoug S M and Bahlouli H 2015  {\em Phys. Rev. A} {\bf 92} 033605
\bibitem{chi14}Chiquillo E 2014 {\em Laser Phys.} {\bf 24} 085502
\bibitem{cue09}Cuevas J, Malomed B A, Kevrekidis P G, and Frantzeskakis D J 2009 {\em Phys. Rev. A} {\bf 79} 053608 
\bibitem{eic12}Eichler R, Zajec D, K\"oberle P, Main J and Wunner G 2012 {\em Phys. Rev. A} {\bf 86} 053611
\bibitem{gao21}Gao P et al 2021 {\em J. Phys. B: At. Mol. Opt. Phys.} {\bf 54} 135301
\bibitem{abd14}Abdullaev F Kh et al 2014 {\em J. Phys. B: At. Mol. Opt. Phys.} {\bf 47} 075301
\bibitem{you11} Young-S L E et al 2011 {\em J. Phys. B: At. Mol. Opt. Phys.} {\bf 44} 101001
\bibitem{edm17}Edmonds M J, Bland T, Doran R and Parker N G 2017 {\em New J. Phys.} {\bf 19} 023019
\bibitem{abd13}Abdullaev F Kh, Gammal A, Malomed B A and Tomio L 2013 {\em Phys. Rev. A} {\bf 87} 063621
\bibitem{pol13}Polo J and Ahufinger V 2013 {\em Phys. Rev. A} {\bf 88} 053628 
\bibitem{hel15}Helm J L, Cornish S L and Gardiner S A 2015 {\em Phys. Rev. Lett.} {\bf 114} 134101
\bibitem{rag15}Raghunandan M, Mishra C, Lakomy K, Pedri P, Santos L and Nath R 2015 {\em Phys. Rev. A} {\bf 92} 013637
\bibitem{nat08}Nath R, Pedri P and Santos L 2008 {\em Phys. Rev. Lett.} {\bf 101} 210402
\bibitem{mis16}Mishra C and Nath R {\em Phys. Rev. A} {\bf 94} 033633
\bibitem{ped05} Pedri P and Santos L 2005 {\em Phys. Rev. Lett.} {\bf 95} 200404
\bibitem{tik08}Tikhonenkov I, Malomed B A, and Vardi A 2008 {\em Phys. Rev. Lett.} {\bf 100} 090406
\bibitem{che17}Chen X Y, Chuang Y L, Lin C Y, Wu C M, Li Y Y, Malomed B A and Lee R K 2017 {\em Phys. Rev. A} {\bf 96} 043631
\bibitem{hua10} Huang C-C and Wu W-C 2010 {\em Phys. Rev. A} {\bf 82} 053612
\bibitem{kla09} Klawunn M and Santos L 2009 {\em Phys. Rev. A} {\bf 80} 013611
\bibitem{kob09}K\"oberle P and Wunner G 2009 {\em Phys. Rev. A} {\bf 80} 063601
\bibitem{lak12c}\L{}akomy K, Nath R and Santos L 2012 {\em Phys. Rev. A} {\bf 86} 013610
\bibitem{lak12b}\L{}akomy K, Nath R and Santos L 2012 {\em Phys. Rev. A} {\bf 86} 023620
\bibitem{mul11}M\"uller S, Billy J, Henn E A L, Kadau H, Griesmaier A, Jona-Lasinio M, Santos L and Pfau T 2011 {\em Phys. Rev. A} {\bf 84} 053601
\bibitem{lak12a}\L{}akomy K, Nath R and Santos L 2012 {\em Phys. Rev. A} {\bf 85} 033618
\bibitem{elh19} Elhadj K M et al 2019 {\em Phys. Scr.} {\bf 94} 085402
\bibitem{cai10} Cai Y, Rosenkranz M, Lei Z and Bao W 2010 {\em Phys. Rev. A} {\bf 82} 043623
\bibitem{wac16}W\"achtler F and Santos L 2016 {\em Phys. Rev. A} {\bf 93} 061603(R)
\bibitem{edl17}Edler D, Mishra C, W\"achtler F, Nath R, Sinha S and Santos L
2017 {\em Phys. Rev. Lett.} {\bf 119} 050403
\bibitem{pri21}Pricoupenko A and Petrov D S 2021 {\em Phys. Rev. A} {\bf 103} 033326
 \end{thebibliography}
\end{document}